\documentclass[letterpaper,english,prl,showpacs,preprint,superscriptaddress]{revtex4}
\usepackage{lmodern}

\usepackage[T1]{fontenc}
\usepackage[latin9]{inputenc}
\usepackage{amsmath}
\usepackage{amssymb}
\usepackage{esint}

\makeatletter

\@ifundefined{textcolor}{}
{%
 \definecolor{BLACK}{gray}{0}
 \definecolor{WHITE}{gray}{1}
 \definecolor{RED}{rgb}{1,0,0}
 \definecolor{GREEN}{rgb}{0,1,0}
 \definecolor{BLUE}{rgb}{0,0,1}
 \definecolor{CYAN}{cmyk}{1,0,0,0}
 \definecolor{MAGENTA}{cmyk}{0,1,0,0}
 \definecolor{YELLOW}{cmyk}{0,0,1,0}
}
\newenvironment{lyxlist}[1]
{\begin{list}{}
{\settowidth{\labelwidth}{#1}
 \setlength{\leftmargin}{\labelwidth}
 \addtolength{\leftmargin}{\labelsep}
 }}
{\end{list}}

\usepackage{amsthm}\usepackage{latexsym}\usepackage{bm}\usepackage{amsfonts}
\makeatother

\usepackage{babel}
\begin{document}

\title{Analytical methods for describing charged particle dynamics in general
focusing lattices using generalized Courant-Snyder theory}

\author{Hong Qin}

\affiliation{Plasma Physics Laboratory, Princeton University, Princeton, NJ 08543}

\affiliation{Department of Modern Physics, University of Science and Technology
of China, Hefei, Anhui 230026, China}

\author{Ronald C. Davidson}

\affiliation{Plasma Physics Laboratory, Princeton University, Princeton, NJ 08543}

\author{Joshua W. Burby}

\affiliation{Plasma Physics Laboratory, Princeton University, Princeton, NJ 08543}

\author{Moses Chung}

\affiliation{Accelerator Physics Center, Fermi National Accelerator Laboratory,
Batavia, IL 60510}
\begin{abstract}
The dynamics of charged particles in general linear focusing lattices
with quadrupole, skew-quadrupole, dipole, and solenoidal components,
as well as torsion of the fiducial orbit and variation of beam energy
is parameterized using a generalized Courant-Snyder (CS) theory, which
extends the original CS theory for one degree of freedom to higher
dimensions. The envelope function is generalized into an envelope
matrix, and the phase advance is generalized into a 4D symplectic
rotation, or an $U(2)$ element. The 1D envelope equation, also known
as the Ermakov-Milne-Pinney equation in quantum mechanics, is generalized
to an envelope matrix equation in higher dimensions. Other components
of the original CS theory, such as the transfer matrix, Twiss functions,
and CS invariant (also known as the Lewis invariant) all have their
counterparts, with remarkably similar expressions, in the generalized
theory. The gauge group structure of the generalized theory is analyzed.
By fixing the gauge freedom with a desired symmetry, the generalized
CS parameterization assumes the form of the modified Iwasawa decomposition,
whose importance in phase space optics and phase space quantum mechanics
has been recently realized. This gauge fixing also symmetrizes the
generalized envelope equation and express the theory using only the
generalized Twiss function $\beta.$ The generalized phase advance
completely determines the spectral and structural stability properties
of a general focusing lattice. For structural stability, the generalized
CS theory enables application of the Krein-Moser theory to greatly
simplify the stability analysis. The generalized CS theory provides
an effective tool to study coupled dynamics and to discover more optimized
lattice design in the larger parameter space of general focusing lattices.
\end{abstract}

\pacs{29.27.-a,52.20.Dq}

\maketitle

\section{Introduction\label{sec:Introduction}}

In accelerators and storage rings, charged particles are confined
transversely by electromagnetic focusing lattices. Many different
kinds of focusing lattice have been successfully designed and implemented.
The fundamental theoretical tool in designing an uncoupled quadrupole
lattice is the Courant-Snyder (CS) theory \cite{Courant58}, which
can be summarized as follows. For a given set of focusing lattice
in the $x-$ and $y-$directions $\kappa_{x}(t)$ and $\kappa_{y}(t)$,
particle's dynamics is governed by the oscillation equation 
\begin{equation}
\ddot{q}+\kappa_{q}(t)q=0\,,\label{q}
\end{equation}
where $q$ represents one of the transverse coordinates, either $x$
or $y.$ Solution of Eq.\,\eqref{q} can be expressed as a symplectic
linear map $M(t)$ that advances the phase space coordinates 
\begin{align}
\binom{q}{\dot{q}} & =M\left(t\right)\binom{q_{0}}{\dot{q}_{0}}\,.
\end{align}
In CS theory, the linear map $M(t)$ is given as 
\begin{equation}
M\left(t\right)=\left(\begin{array}{cc}
\sqrt{\dfrac{\beta}{\beta_{0}}}\left[\cos\phi+\alpha_{0}\sin\phi\right] & \sqrt{\beta\beta_{0}}\sin\phi\\
-\dfrac{1+\alpha\alpha_{0}}{\sqrt{\beta\beta_{0}}}\sin\phi+\dfrac{\alpha_{0}-\alpha}{\sqrt{\beta\beta_{0}}}\cos\phi & \sqrt{\dfrac{\beta_{0}}{\beta}}\left[\cos\phi-\alpha\sin\phi\right]
\end{array}\right)\,,\label{eq:M1}
\end{equation}
where $\alpha\left(t\right)$ and $\beta\left(t\right)$ are two of
the so-called Twiss parameters, and $\phi\left(t\right)$ is the phase
advance. They are defined by an envelope function $w\left(t\right)$
as
\begin{align}
\beta\left(t\right) & =w^{2}\left(t\right)\,,\label{beta}\\
\alpha\left(t\right) & =-w\dot{w}\,,\label{eq:alpha}\\
\phi\left(t\right) & =\int_{0}^{t}\dfrac{dt}{\beta\left(t\right)}\,,\label{phi}
\end{align}
and the envelope function $w\left(t\right)$ is determined by the
envelope equation 
\begin{equation}
\ddot{w}+\kappa_{q}\left(t\right)w=w^{-3}\,.\label{w1}
\end{equation}
In Eq.\,\eqref{eq:M1}, $q_{0}=q\left(t=0\right),$ $\dot{q}_{0}=\dot{q}\left(t=0\right),$
$\beta_{0}=\beta\left(t=0\right),$ and $\alpha_{0}=\alpha\left(t=0\right)$
are the initial conditions at $t=0.$ 

Associated with the dynamics of Eq.\,\eqref{q}, there exists a constant
of motion, $I_{CS},$ known as the Courant-Synder invariant
\begin{gather}
I_{CS}=\dfrac{q^{2}}{w^{2}}+\left(w\dot{q}-\dot{w}q\right)^{2}=(q,\dot{q})\left(\begin{array}{cc}
\gamma & \alpha\\
\alpha & \beta
\end{array}\right)\left(\begin{array}{c}
q\\
\dot{q}
\end{array}\right)\,,\label{csi}\\
\gamma(t)\equiv w^{-2}+\dot{w}^{2}\,.\label{gamma}
\end{gather}
Here $\gamma\left(t\right)$ is the third Twiss parameter. It turns
out that the transfer matrix $M\left(t\right)$ can be decomposed
into the elegant form \cite{Lee99-47}
\begin{equation}
M\left(t\right)=\left(\begin{array}{cc}
w & 0\\
\dot{w} & \dfrac{1}{w}
\end{array}\right)\left(\begin{array}{cc}
\cos\phi & \sin\phi\\
-\sin\phi & \cos\phi
\end{array}\right)\left(\begin{array}{cc}
w_{0}^{-1} & 0\\
-\dot{w}_{0} & w_{0}
\end{array}\right)\,,\label{m1d}
\end{equation}
which seems to indicate a certain structure for $M(t)$. 

The CS theory can be viewed as a parameterization method of the time-dependent
$2\times2$ symplectic matrix $M(t)$ for a standard uncoupled lattice.
Not surprisingly, there exist other parameterization schemes mathematically.
Why is the CS parameterization preferable? This is because it describes
the physics of charged particle dynamics. The main components of the
CS theory\emph{,} i.e., the phase advance, the envelope equation,
the transfer matrix, and the CS invariant are physical quantities
describing the dynamics of the particles. For example, the CS invariant
defines the emittance in phase space, and the envelope function describes
the transverse dimensions in configuration space. This theoretical
framework also makes it possible to investigate collective effects
associated with high-intensity beams, such as in the construction
of the Kapchinskij-Vladimirskij distribution \cite{KV59,Qin09-PRL,Qin13PRL}. 

However, the CS theory can only be applied to the $x-$ or $y-$dynamics
separately for the ideal case of uncoupled quadrupole focusing lattices.
In realistic accelerators, there exist bending magnets, torsion of
the design orbit (fiducial orbit), and skew-quadrupole components,
which are introduced intentionally or by misalignment \cite{Barnard96,Kishek99}.
Solenoidal magnets are also used in certain applications \cite{Friedman10}.
When these additional components are included, the transverse dynamics
in the $x-$ and $y-$directions are coupled, and the focusing force
depends on the transverse momentum as well. In this most general case,
the transfer matrix $M(t)$ is a time-dependent $4\times4$ symplectic
matrix, which has $10$ time-dependent parameters and admits many
different schemes for parameterization. The first set of parameterization
schemes for $M\left(t\right)$ were developed by Teng and Edwards
\cite{Teng71,Edwards73,Teng03} and Ripken \cite{Ripken70,Wiedemann07-614,Lebedev10},
some of which have been adopted in lattice design and particle tracking
codes, such as the MAD code \cite{Grote89,DragtMM}. A class of different
parameterizations by directly generalizing the Twiss parameters to
higher dimensions has also been developed by Dattoli, et al. \cite{Dattoli92,Dattoli92b,Dattoli92c}.
However, in contrast to the original CS theory, these parameterization
schemes are designed from mathematical considerations, and fail to
connect with physical parameters of the beam. The elegant and much-needed
connection with the physics of beam dynamics in the original CS theory
for one degree of freedom is not transparent in these parameterization
schemes. This is probably why there is no \textit{de facto} standard
yet adopted by the accelerator community. Another main reason is that
for most present-day accelerators and rings, the transverse dynamics
are so nearly decoupled that perturbative treatment often works satisfactorily.
Even for lattices with strong coupling, elementary methods can be
used to analyze the dynamics, even though the calculation often becomes
rather involved and requires diligence and patience.

In a recent Letter \cite{Qin13PRL2}, we reported the development
of a generalized CS theory for focusing lattices with the most general
form in Eq.\,\eqref{H}, including bending magnets, torsion of the
design orbit, and solenoidal magnets, in addition to quadrupole and
skew-quadrupole components. In this generalized theory, the physics
elements of the original CS theory, i.e., the phase advance, the envelope
equation, the transfer matrix, and the CS invariant are all generalized
to the 2D coupled case with identical structure. This new development
also generalizes our previous results for coupled dynamics including
only a skew-quadrupole lattice component \cite{Qin09-NA,Qin09PoP-NA,Chung10,Qin11-056708}.
In this paper, we give a detailed derivation of the generalized CS
theory reported in Ref. \cite{Qin13PRL2}, describe the theoretical
structure of the theory in terms of gauge freedoms and group decomposition,
and demonstrate the application of the theory in stability analysis.

\section{Theoretical model and summary of results\label{sec:Theoretical-model}}

In this section, we outline the theoretical methods used and summarize
the main results obtained in this paper. As discussed in Sec. \ref{sec:Introduction},
when realistic components such as skew-quadrupoles, bending magnets,
torsion of the design orbit, solenoidal magnets are included, in addition
to the standard quadrupole components, the transverse dynamics in
the $x-$ and $y-$directions are coupled, and the focusing force
depends on the transverse momentum. In this case, the linear dynamics
of a charged particle relative to the fiducial orbit are governed
by a general time-dependent Hamiltonian \cite{Micheloetti95-166}
of the form 
\begin{align}
H & =\frac{1}{2}z^{T}Az\,,\,\,\,A=\left(\begin{array}{cc}
\kappa\left(t\right) & R\left(t\right)\\
R\left(t\right)^{T} & m^{-1}\left(t\right)
\end{array}\right)\,.\label{H}
\end{align}
Here, $z=\left(x,y,p_{x},p_{y}\right)^{T}$ are the transverse phase
space coordinates, and $\kappa(t),$ $R\left(t\right)$ and $m^{-1}\left(t\right)$
are time-dependent $2\times2$ matrices. The matrices $A$, $\kappa(t)$
and $m^{-1}\left(t\right)$ are also symmetric. In this most general
Hamiltonian, the skew-quadrupole and dipole components are included
in the off-diagonal terms of the $\kappa\left(t\right)$ matrix, and
the solenoidal component and the torsion of the fiducial orbit are
included in the $R\left(t\right)$ matrix. There are several different
methods to include the effect of torsion, which were reviewed by Hoffstaetter
\cite{Hoffstaetter95}. Typically, Frenet-Serret coordinates along
the fiducial orbit are used. When the fiducial orbit is straight,
the Frenet-Serret coordinates are not uniquely defined. In this case,
we can choose any particular set of Frenet-Serret coordinates in the
straight section, as long it is smoothly connected to those in the
curved sections. The variation of beam energy along the fiducial orbit
is reflected in the mass matrix $m^{-1}\left(t\right),$ which is
allowed to be any real symmetric matrix for complete generality. The
transfer matrix $M(t)$ corresponding to $H$ is a time-dependent
$4\times4$ symplectic matrix, which has $10$ time-dependent parameters.
Our goal is to develop a generalized Courant-Snyder parameterization
method for $M(t),$ which has the same elegant structure and direct
connection to beam dynamics as the original Courant-Snyder theory
for one degree of freedom. 

We will use a time-dependent symplectic transformation technique \cite{Leach77,Qin09-NA,Qin11-056708}
to analyze the charged particle dynamics governed by the Hamiltonian
given in Eq.\,\eqref{H}. This technique is described in Sec. \ref{sec:Time-dependent}.
The concept of scalar envelope function is generalized to a $2\times2$
envelope matrix, and the envelope equation in $2\times2$ matrix form
is developed {[}see Eq.\,\eqref{w}{]}. In the original CS theory,
the envelope equation \eqref{w1} is one dimensional and plays a central
role. It also has been discovered or re-discovered many times \cite{Ermakov80,Milne30,Pinney50,Lewis68,Lewis69}
in other branches of physics. In quantum physics, it is known as the
Ermakov-Milne-Pinney equation \cite{Ermakov80,Milne30,Pinney50},
which has been utilized to study 1D time-dependent quantum systems
\cite{Morales88,Monteoliva94} and associated non-adiabatic Berry
phases \cite{Berry85}. A brief account of the history of the 1D envelope
equation can be found in Ref. \cite{Qin06Sym}. We expect the generalization
of the envelope equation to higher dimensions for the most general
Hamiltonian to have applications in areas other than beam physics
as well. The 1D CS invariant given by Eq.\,\eqref{csi}, also known
as the Lewis invariant \cite{Lewis68,Lewis69} in quantum physics,
is generalized to higher dimensions in Eqs.\,\eqref{Ixi} and \eqref{eq:ICS2}. 

Also in Sec.\,\ref{sec:Generalized-Courant-Snyder}, the 1D phase
advance is generalized to a time-dependent matrix $P$, which belongs
to the symplectic rotation group $Sp(4)\bigcap SO(4)=U(2).$ Here,
$Sp(4)$, $SO(4),$ and $U(2)$ denote the groups of $4\times4$ symplectic
matrices, $4\times4$ rotation matrices, and $2\times2$ unitary matrices,
respectively. For dynamics with one degree of freedom, the phase advance
is naturally an angle (an element of $SO(2)$) in the 2D phase space.
For dynamics with two degrees of freedom, the phase space is 4D, and
it is tempting to represent the phase advance by two angles. This
is what has been adopted in previous parameterization schemes. From
the viewpoint of theoretical physics and geometry, however, it is
more natural to represent the phase advance for dynamics with two
degrees of freedom by a 4D rotation (an element of $SO(4)$), which
is not equivalent to two 2D rotations. Because of the symplectic nature
of the Hamiltonian dynamics, the generalized phase advance in higher
dimensions thus belongs to the symplectic rotation group. Of course,
one can adopt different views on this. In the normal form analysis
of accelerator rings, the 4D transfer matrix is reduced to a 2D rotation
after block diagonalization. In a sense, we can compare coupled betatron
motion to the Dirac equation. In a fully quantum mechanical limit,
the only correct approach is to treat electrons and positrons as inextricably
coupled; only the 4D approach is permissible in this limit. But in
the accelerator physics, we routinely treat electrons and positrons
as completely separate entities, and two 2D descriptions are adopted
without much hesitation. 

The generalized decomposition for the symplectic map $M(t)$ is given
by Eq.\,\eqref{Md}, which has exactly the same structure as the
original 1D CS theory given by Eq.\,\eqref{m1d}. In addition to
its aesthetic elegance, the generalized CS theory provides an effective
tool to describe the beam dynamics governed by the most general Hamiltonian.
The $2\times2$ envelope matrix $w$ defines the transverse dimension
of the beam, and the generalized CS invariant defines the emittance.
These components of the generalized CS theory are derived in detail
in Sec. \ref{sec:Generalized-Courant-Snyder}. For the present application
to beam transverse dynamics, there are two degrees of freedom. But
the theory developed is valid for any degree of freedom. For a system
with $n$-degrees of freedom, the time-dependent matrix $A(t)$ specifying
the Hamiltonian in Eq.\,\eqref{H} will be $2n\times2n,$ the envelope
matrix will be $n\times n$, and the phase advance will belong to
$Sp(2n)\bigcap SO(2n)=U(n).$ 

In Sec. \ref{sec:Group-structure}, we investigate the group structure
of the generalized CS theory, which is built on the decomposition
of the time-dependent symplectic coordinate transformation $G$ in
the form of Eq.\,\eqref{G}. There exists a gauge freedom in this
decomposition specified by a 2D rotation element $c\in SO(2)$ for
every $t$. The transfer map $M(t)$ is independent of this gauge.
By fixing the gauge freedom with a desired symmetry, the decomposition
of $G$ as $PS$ assumes the form of the modified Iwasawa decomposition
(or pre-Iwasawa decomposition), whose importance in phase space optics
\cite{Simon98,Wolf04-173} and phase space quantum mechanics \cite{deGosson06-42}
has been recently realized. This specific gauge fixing also symmetrizes
the generalized envelope equation and express the theory using only
the generalized Twiss function $\beta.$ For a symplectic matrix,
the modified Iwasawa decomposition is equivalent to the well-known
Iwasawa decomposition for a semi-simple Lie group \cite{Iwasawa49}.
However, the unique feature of the theory described here is that the
decomposition is constructed as a function of time, and from the viewpoint
of dynamics using the generalized envelope equation. Nevertheless,
it is a pleasant surprise to find the deep connection between the
original CS theory for charged particle dynamics \cite{Courant58}
and the Iwasawa decomposition for Lie groups \cite{Iwasawa49}, two
theoretical formalisms developed concurrently. This connection also
demonstrates that beam dynamics, phase space optics and quantum dynamics
have a similar theoretical structure at the fundamental level. In
order to satisfy the symmetry requirement of the modified Iwasawa
decomposition, the gauge freedom need to be selected locally as a
function of time, which is the characteristics of gauge theories in
theoretical physics. This procedure also results in a symmetrized
envelope equation in terms of the generalized Twiss parameter $\beta$,
which is a symmetric, positive-definite matrix. The beam dimensions
and emittance can be expressed using the $\beta$ matrix only. 

We show in Sec. \ref{sec:Stability-analysis} how the generalized
CS theory can be used to analyze the stability of a charge particle
dynamics in realistic accelerators with quadrupole, skew-quadrupole,
dipole, and solenoidal components, as well as torsion of the fiducial
orbit and variation of beam energy. It turns out that the generalized
phase advance as a symplectic rotation completely determines the spectral
and structural stability properties of the general lattice after a
matched solution of the envelope equation is found. For structural
stability, the generalized CS theory enables us to apply the Krein-Moser
theory \cite{Krein50,Gelfand55,Moser58,Yakubovich75,DragtBook} to
greatly simplify the stability analysis. This general result includes
the well-known stability criterion for sum/difference resonances for
uncoupled quadrupole lattices as a special case.

\section{Method of time-dependent canonical transformation\label{sec:Time-dependent}}

We will construct the generalized Courant-Snyder theory for the general
focusing lattice given by Eq.\,\eqref{H} using a method of time-dependent
canonical coordinate transformation. Let's consider a linear, time-dependent
Hamiltonian system with n-degrees of freedom
\begin{align}
H & =\frac{1}{2}z^{T}A\left(t\right)z\,,\label{Hn}\\
z & =\left(x_{1},x_{2},...,x_{n},p_{1},p_{2},...,p_{n}\right)^{T}\,.\nonumber 
\end{align}
 Here, $A\left(t\right)$ is a $2n\times2n$ time-dependent, symmetric
matrix. The Hamiltonian in Eq.\,(\eqref{H}) has this form with $n=2.$
The basic idea is to introduce a time-dependent linear canonical transformation
\cite{Leach77} 
\begin{equation}
\bar{z}=S\left(t\right)z\,,\label{zsz}
\end{equation}
 such that in the new coordinates $\bar{z},$ the transformed Hamiltonian
has the desired form
\begin{equation}
\bar{H}=\frac{1}{2}\bar{z}^{T}\bar{A}\left(t\right)\bar{z}\,,
\end{equation}
where $\bar{A}\left(t\right)$ is a targeted symmetric matrix. Because
the transformation \eqref{zsz} is canonical, it requires that 
\begin{equation}
SJS^{T}=J\,,\label{eq:sjsj}
\end{equation}
Here, $J$ the $2n\times2n$ unit symplectic matrix of order $2n$,
\begin{equation}
J=\left(\begin{array}{cc}
0 & I\\
-I & 0
\end{array}\right)\,,
\end{equation}
and $I$ is the $n\times n$ unit matrix. Equation \eqref{eq:sjsj}
implies that $S$ is a symplectic matrix. In addition, it needs to
satisfy a differential equation, which can be derived as follows.
Hamilton's equation for $z$ is given by
\begin{align}
\dot{z} & =J\nabla H\,,\label{eq:zdot}
\end{align}
Using index notation, Eq.\,\eqref{eq:zdot} becomes 
\begin{align}
\dot{z}_{j} & =J_{ij}\dfrac{\partial H}{\partial z_{j}}=\frac{1}{2}J_{ij}\left(\delta_{lj}A_{lm}z_{m}+z_{l}A_{lk}\delta_{kj}\right)\nonumber \\
 & =\frac{1}{2}J_{ij}\left(A_{jm}+A_{mj}\right)z_{m}=JA_{jm}z_{m}\,.\label{zdot-ind}
\end{align}
Switching back to matrix notation, Eq.\,(\ref{zdot-ind}) can be
expressed as 
\begin{equation}
\dot{z}=JAz\,.\label{zd}
\end{equation}
Similarly, 
\begin{equation}
\dot{\bar{z}}=J\bar{A}\bar{z}=J\bar{A}Sz\,.\label{zbd}
\end{equation}
Meanwhile, $\dot{\bar{z}}$ can be directly calculated from Eq.\,(\ref{zsz})
by taking a time-derivative, which gives 
\begin{equation}
\dot{\bar{z}}=\dot{S}z+S\dot{z}=\left(\dot{S}+SJA\right)z\,.\label{zbd2}
\end{equation}
 Combining Eqs.\,(\ref{zbd}) and (\ref{zbd2}) gives the differential
equation for $S$
\begin{equation}
\dot{S}=\left(J\bar{A}S-SJA\right)\,.\label{S}
\end{equation}

The remarkable feature of the canonical transformation $S$ is that
it is always symplectic, if $S$ is initially symplectic at $t=0$.
This assertion can be proved by two methods. For the first proof,
we follow Leach \cite{Leach77} and consider the dynamics of the matrix
$K=SJS^{T},$ 
\begin{align}
\dot{K} & =\dot{S}JS^{T}+SJ\dot{S}^{T}\nonumber \\
 & =\left[\left(J\bar{A}S-SJA\right)JS^{T}+SJ\left(-S\bar{A}J+AJS^{T}\right)\right]\nonumber \\
 & =\left[J\bar{A}SJS^{T}-SJS^{T}\bar{A}J\right]=\left[J\bar{A}K-K\bar{A}J\right]\,.\label{Kdot}
\end{align}
Equation (\ref{Kdot}) has a fixed point at $K=J.$ If $S(t=0)$ is
symplectic, \textit{i.e.}, $K\left(t=0\right)=J,$ then $\dot{K}=0$
and $K=J$ for all $t$, and $S$ is symplectic for all $t.$ A more
geometric proof can be given from the viewpoint of the flow of $S$.
Because $A$ is symmetric, we have $JJ\bar{A}-\bar{A}^{T}JJ=0,$ which
indicates that $J\bar{A}$ belongs to the Lie algebra $sp\left(2n,R\right).$
We now show that if $S$ is symplectic at a given $t,$ then $J\bar{A}S$
belongs to the tangent space of $Sp\left(2n,R\right)$ at $S$, i.e.,
$J\bar{A}S\in T_{S}SP\left(2n,R\right).$ Let's examine the Lie group
right action: $S:$ $a$ $\mapsto aS$ for any $a$ in $Sp\left(2n,R\right),$
and the associated tangent map 
\begin{equation}
T_{S}:\ T_{a}Sp\left(2n,R\right)\rightarrow T_{aS}Sp\left(2n,R\right).
\end{equation}
It is evident that $J\bar{A}S$ is the image of the Lie algebra element
$J\bar{A}$ under the tangential map $T_{S}.$ This means that $J\bar{A}S$
is a vector\ tangential to the space of $Sp\left(2n,R\right)$ at
$S,$ if $S$ is on $Sp\left(2n,R\right)$. By the same argument $SJA\in T_{S}SP\left(2n,R\right)$
as well. Thus, the right-hand side of Eq.\,\eqref{S} is a vector
on $Sp\left(2n,R\right)$, and the $S$ dynamics will stay on the
space of $Sp(2n,R)$. We can always choose initial conditions such
that $S$ is symplectic at $t=0,$ and this will guarantee that the
time-dependent transformation specified by Eq.\,(\ref{S}) is symplectic
for all $t.$

\section{Generalized Courant-Snyder theory\label{sec:Generalized-Courant-Snyder}}

We now apply the technique developed in Sec. \ref{sec:Time-dependent}
to the Hamiltonian system in Eq.\,\eqref{H}. Our goal is to find
a new coordinate system where the transformed Hamiltonian vanishes.
This idea is identical to that in Hamilton-Jacobi theory. Applications
of Hamilton-Jacobi theory include the construction of action-angle
variables for periodic systems \cite{Goldstein80-456} and finding
geodesic curves on an ellipsoid \cite{Arnold89-261}. It is often
required that the variables in the Hamilton-Jocabi equation can be
separated in order for the technique to be effective for practical
problems. This limits its application. Since our dynamics is linear,
the new coordinate system can be more easily constructed using the
method developed in Sec. \ref{sec:Time-dependent}. We will accomplish
this goal in two steps. First, we seek a coordinate transformation
$\bar{z}=Sz$ such that, in the $\bar{z}$ coordinates, the Hamiltonian
assumes the form 
\begin{equation}
\bar{H}=\frac{1}{2}\bar{z}^{T}\bar{A}\bar{z}\,,\,\,\bar{A}=\left(\begin{array}{cc}
\mu(t) & 0\\
0 & \mu(t)
\end{array}\right)\,,
\end{equation}
where $\mu(t)$ is a $2\times2$ matrix to be determined. To write
Eq.\,\eqref{S} in the format of $2\times2$ blocks, we let 
\[
S=\left(\begin{array}{cc}
S_{1} & S_{2}\\
S_{3} & S_{4}
\end{array}\right),
\]
and split the differential equation for $S$, i.e., Eq.\,\eqref{S},
into four matrix equations, 
\begin{align}
\dot{S}_{1} & =\mu S_{3}-S_{1}R^{T}+S_{2}\kappa\,\,,\label{eq:S1-1}\\
\dot{S}_{2} & =\mu S_{4}-S_{1}m^{-1}+S_{2}R\,,\label{eq:S2-1}\\
\dot{S}_{3} & =-\mu S_{1}-S_{3}R^{T}+S_{4}\kappa\,,\\
\dot{S}_{4} & =-\mu S_{2}-S_{3}m^{-1}+S_{4}R\,.\label{eq:S4-1}
\end{align}
Including $\mu(t),$ we have five $2\times2$ matrices unknown. The
extra freedom is introduced by the to-be-determined $\mu(t)$. Based
on the analogy with Eq.\,\eqref{m1d}, we choose $S_{2}\equiv0$
to remove the freedom. We rename $S_{4}$ to be $w,$ i.e., $w\equiv S_{4}$,
because it will be clear later that $S_{4}$ is the envelope matrix.
Equations \eqref{eq:S1-1}-\eqref{eq:S4-1} become
\begin{align}
\dot{S}_{1} & =\mu S_{3}-S_{1}R^{T}\,,\label{S1}\\
S_{1} & =\mu wm\,,\label{S2}\\
\dot{S}_{3} & =-\mu S_{1}-S_{3}R^{T}+w\kappa\,,\label{S3}\\
S_{3} & =-\dot{w}m+wRm,\label{S4}
\end{align}
for matrices $S_{1}$, $S_{3}$, $w$ and $\mu$. Because $(S_{1},S_{2}=0,S_{3},S_{4}=w)$
describes a curve in $Sp(4),$ they are consistent with the symplectic
condition $S_{1}S_{4}^{T}-S_{2}S_{3}^{T}=I,$ i.e., $S_{1}w^{T}=I,$
which implies
\begin{equation}
S_{1}=w^{-T}\,.\label{eq:S1}
\end{equation}
From Eq.\,\eqref{S2}, we obtain 
\begin{equation}
\mu=\left(wmw^{T}\right)^{-1}.\label{mu}
\end{equation}
It is straightforward to verify that Eq.\,\eqref{S1} is equivalent
to another symplectic condition $S_{3}S_{4}^{T}=S_{4}S_{3}^{T}.$
Substituting Eqs.\,\eqref{S3}-\eqref{mu} into Eq.\,\eqref{S3},
we immediately obtain the following matrix differential equation for
the envelope matrix $w,$
\begin{equation}
\frac{d}{dt}\left(\frac{dw}{dt}m-wRm\right)+\frac{dw}{dt}mR^{T}+w\left(\kappa-RmR^{T}\right)-\left(w^{T}wmw^{T}\right)^{-1}=0\,.\label{w}
\end{equation}
This is the desired generalized envelope equation. It generalizes
the 1D envelope equation \eqref{w1}, or the Ermakov-Milne-Pinney
equation \cite{Ermakov80,Milne30,Pinney50}, as well as the previous
matrix envelope equation for cases with only quadrupole and skew-quadrupole
magnets, i.e., $R=0$ \cite{Qin09-NA,Qin09PoP-NA,Chung10,Qin11-056708}.
For $n$-degrees of freedom, the envelope matrix $w$ will be $n\times n$,
and the generalized envelope equation has the same form as Eq.\,\eqref{w}. 

Once $w$ is solved for from the envelope equation, we can determine
$S_{1}$ from Eq.\,\eqref{eq:S1} and $S_{3}$ from Eq.\,\eqref{S4}.
In terms of the envelope matrix $w,$ the symplectic transformation
$S$ and its inverse are given by
\begin{align}
S & =\left(\begin{array}{cc}
w^{-T} & 0\\
(wR-\dot{w})m & w
\end{array}\right)\,,\label{eq:S=00003D}\\
S^{-1} & =\left(\begin{array}{cc}
w^{T} & 0\\
\left(w^{-1}\dot{w}-R\right)mw^{T} & w^{-1}
\end{array}\right)\,.\label{eq:S-1=00003D}
\end{align}

The second step is to use another coordinate transformation $\bar{\bar{z}}=P(t)\bar{z}\,$
to transform $\bar{H}$ into a vanishing Hamiltonian $\bar{\bar{H}}\equiv0$
at all time, thereby rendering the dynamics trivial in the new coordinates.
The determining equation for the transformation $P(t)$ is 
\begin{equation}
\dot{P}=-PJ\bar{A}=P\left(\begin{array}{cc}
0 & -\mu\\
\mu & 0
\end{array}\right)\,.\label{P}
\end{equation}
As explained in Sec. \ref{sec:Time-dependent}, the $P$ matrix satisfying
Eq.\,\eqref{P} is symplectic because $J\bar{A}\in sp(4).$ From
$\mu=\mu^{T}$, we know that $J\bar{A}$ is also antisymmetric, i.e.,
$J\bar{A}$ belongs to the Lie algebra $so(4)$ of the 4D rotation
group $SO(4)$. Thus $J\bar{A}\in sp(4)\bigcap so(4)$, and $P(t)$
is a curve in the group of $4$D symplectic rotations, i.e., $P(t)\in Sp(4)\bigcap SO(4)=U(2)$,
provided the initial condition of $P(t)$ is chosen such that $P(0)\in Sp(4)\bigcap SO(4)=U(2)$.
We call $P(t)$ the generalized phase advance, an appropriate descriptor
in light of the fact that $P(t)$ is a symplectic rotation. The Lie
algebra element (infinitesimal generator) $-J\bar{A}=\left(\begin{array}{cc}
0 & -\mu\\
\mu & 0
\end{array}\right)$ is the phase advance rate, and it is determined by the envelope matrix
through Eq.\,\eqref{mu}. Since $Sp(4)\bigcap SO(4)=U(2)$, $P$
and its inverse must have the forms
\begin{align}
P & =\left(\begin{array}{cc}
P_{1} & P_{2}\\
-P_{2} & P_{1}
\end{array}\right)\,,\label{eq:P=00003D}\\
P^{-1} & =P^{T}=\left(\begin{array}{cc}
P_{1}^{T} & -P_{2}^{T}\\
P_{2}^{T} & P_{1}^{T}
\end{array}\right)\,.\,\,\label{eq:P-1=00003D}
\end{align}

Combining the two symplectic coordinate transformations, we obtain
the transformation 
\begin{equation}
\bar{\bar{z}}=G(t)z=P(t)S(t)z\,.\label{G}
\end{equation}
In the $\bar{\bar{z}}$ coordinate representation, because $\bar{\bar{H}}\equiv0$,
the dynamics is trivial, i.e., $\bar{\bar{z}}=const.$ This enables
us to construct the symplectic matrix specifying the map between $z_{0}$
and $z=M(t)z_{0}$ as
\begin{equation}
M(t)=S^{-1}P^{-1}P_{0}S_{0}=\left(\begin{array}{cc}
w^{T} & 0\\
\left(w^{-1}\dot{w}-R\right)mw^{T} & w^{-1}
\end{array}\right)P^{T}\left(\begin{array}{cc}
w^{-T} & 0\\
(wR-\dot{w})m & w
\end{array}\right)_{0},\label{Md}
\end{equation}
where subscript ``0'' denotes initial conditions at $t=0$, and
$P_{0}$ is taken to be $I$ without loss of generality. This expression
for $M(t)$ generalizes the decomposition of the symplectic map for
the original 1D CS theory given by Eq.\,\eqref{m1d}. The first and
the third matrices in Eq.\,\eqref{m1d} obviously have the same construction
as their counterparts in Eq.\,\eqref{Md}. The phase advance, as
a 4D symplectic rotation $P^{T}$ in Eq.\,\eqref{Md}, generalizes
the 2D rotation matrix, which is also symplectic, in Eq.\,\eqref{m1d}.
The phase advance $P$ is generated by its infinitesimal generator
$J\bar{A}$ determined by the envelope matrix through $\mu=\left(wmw^{T}\right)^{-1}.$
This mechanism for phase advance in 4D phase space is identical to
the original 1D CS theory where the infinitesimal generator of the
phase advance is $w^{-2}$ for a scalar envelope $w.$ The importance
of the decomposition in Eqs.\,\eqref{Md} and \eqref{m1d} can be
appreciated from both physical and mathematical points of view. We
explain the physical meaning of the decomposition here, and leave
the mathematical analysis to Sec. \ref{sec:Group-structure}. The
first matrix from the right is a matching transformation at $t=0$
of the initial conditions to an equivalent focusing system, where
the phase space dynamics can be characterized by a time-dependent
rotation. The second matrix from the right is a transformation along
the time axis in this equivalent focusing system, with the phase advance
playing the role of a time-like evolution parameter. And the third
matrix from the right is a back-transformation to the original coordinate
system at $t>0$.

The coordinate transformation can also be used to construct invariants
of the dynamics. A general description of linear symplectic invariants
can be found in Refs. \cite{Neri90,Dragt92}. For any constant $4\times4$
positive-definite matrix $\xi,$ the quantity 
\begin{equation}
I_{\xi}=z^{T}S^{T}P^{T}\xi PSz\label{Ixi}
\end{equation}
is a constant of motion, since $\bar{\bar{z}}=PSz$ is a constant
of motion. The subscript ``$\xi$'' in $I_{\xi}$ is used to indicate
that it is an invariant associated with $\xi.$ For the special case
of $\xi=I,$ the phase advance $P$ in Eq.\,\eqref{Ixi} drops out,
and 
\begin{equation}
I_{CS}\equiv z^{T}S^{T}Sz=z^{T}\left(\begin{array}{cc}
\gamma & \alpha\\
\alpha^{T} & \beta
\end{array}\right)z\,,\label{eq:ICS2}
\end{equation}
where $\alpha,$ $\beta,$ and $\gamma$ are $2\times2$ matrices
defined by
\begin{align}
\alpha & \equiv w^{T}S_{3}\,,\label{eq:alphag}\\
\beta & \equiv w^{T}w\,,\label{eq:betag}\\
\gamma & \equiv S_{3}^{T}S_{3}+w^{-1}w^{-T}\,.\label{eq:gammag}
\end{align}
Here, we have used $I_{CS}$ to denote this special invariant because
it is the invariant that generalizes the CS invariant \cite{Courant58}
(or Lewis invariant \cite{Lewis68,Lewis69}) for one degree of freedom
in Eq.\,\eqref{csi}. The matrices, $\alpha,$ $\beta,$ and $\gamma$
are the generalized Twiss parameters in higher dimensions. It is straightforward
to verify that they satisfy 
\begin{equation}
\beta\gamma=I+\alpha^{2},\label{bg}
\end{equation}
which is a familiar relationship in the original CS theory between
the scalar Twiss parameters defined by Eqs.\,\eqref{beta}, \eqref{eq:alpha}
and \eqref{gamma}. The symplectic condition $wS_{3}^{T}=S_{3}w^{T}$
has been used in obtaining Eq.\,\eqref{bg}. 

It has been demonstrated that the envelope matrix $w$ and the invariant
$I_{\xi}$ define the beam dimensions and emittance for both low intensity
beams and high intensity beams with strong space-charge potential
\cite{Qin09-PRL,Qin13PRL}. 

Note that we have ``overloaded'' the symbols ``$M$, $w,$ $\alpha,$
$\beta,$ $\gamma,$ $I_{CS}$'' to represent the same physical quantities
in both the original CS theory for one degree of freedom and the generalized
CS theory in higher dimensions without causing any confusion. It is
actually more appropriate to do so than not, because the quantities
in higher dimensions recover their counterparts for one degree of
freedom as special cases, and the correspondence between them is exact.

\section{Group structure of the generalized Courant-Snyder theory -- rotation
gauge and modified Iwasawa decomposition\label{sec:Group-structure}}

In Sec. \ref{sec:Generalized-Courant-Snyder}, we noted that initial
conditions for the envelope matrix $w$ need to satisfy the symplectic
condition; otherwise they can be arbitrary. There are freedoms in
the initial conditions and thus the solutions for $w.$ But the transfer
matrix $M$ is independent from these freedoms, which are thus gauge
freedoms. A subset of the gauge freedoms has the structure of the
orthogonal group $O(n).$ For a time-independent element $c\in O(n),$
we define the gauge transformation $c:\,(w,P)\mapsto(\tilde{w},\tilde{P})$
as 
\begin{align}
\tilde{w} & =cw\,,\label{eq:wt}\\
\tilde{P} & =P\left(\begin{array}{cc}
c^{-1} & 0\\
0 & c^{-1}
\end{array}\right)\,.\label{eq:Pt}
\end{align}
Let's show that the transformed $\tilde{w}$ and $\tilde{P}$ also
satisfy Eqs.\,\eqref{w} and \eqref{P}, respectively, and $M$ is
gauge invariant. Multiplying Eq.\,\eqref{w} by $c$ from the left
to obtain the governing equation for $\tilde{w}$, 
\begin{equation}
\frac{d}{dt}\left(\frac{d\tilde{w}}{dt}m-\tilde{w}Rm\right)+\frac{d\tilde{w}}{dt}mR^{T}+\tilde{w}\left(\kappa-RmR^{T}\right)-\left(\tilde{w}^{T}\tilde{w}m\tilde{w}^{T}\right)^{-1}=0\,,
\end{equation}
which is the same as Eq.\,\eqref{w} with $w$ replaced by $\tilde{w}$.
According to Eq.\,\eqref{mu}, the $\mu$ matrix transforms as
\[
\tilde{\mu}=c\mu c^{-1}\,.
\]
Equation \eqref{P} thus can be re-expressed in the same form using
$\tilde{P}$ and $\tilde{\mu}$ as 
\begin{equation}
\frac{d}{dt}\tilde{P}=\tilde{P}\left(\begin{array}{cc}
0 & -\tilde{\mu}\\
\tilde{\mu} & 0
\end{array}\right)\,.\label{P-1}
\end{equation}
From Eqs.\,\eqref{eq:S=00003D} and \eqref{eq:S-1=00003D}, the $S$
matrix and its inverse transform as
\begin{align}
\tilde{S} & =\left(\begin{array}{cc}
c & 0\\
0 & c
\end{array}\right)S\,,\label{eq:St}\\
\tilde{S}^{-1} & =S^{-1}\left(\begin{array}{cc}
c^{-1} & 0\\
0 & c^{-1}
\end{array}\right)\,.\label{eq:St-1}
\end{align}
Combining Eqs.\, \eqref{eq:wt} and \eqref{eq:Pt}, \eqref{eq:St}
and \eqref{eq:St-1}, we conclude that $M$ is invariant under the
gauge transformation $c:\,(w,P)\mapsto(\tilde{w},\tilde{P})$, i.e.,
$\tilde{M}=M$. 

The $O(n)$ gauge group introduces an equivalent class $([w],[P])$
for the decomposition of $M$ using $(w,P).$ The dimension of this
equivalent class is the dimension of $M$ as a $Sp(2n)$ group. To
specify $S$ by $w$ and $\dot{w}$, $2n^{2}$ numbers are needed.
To specify $P\in U(n),$ additional $n^{2}$ numbers are needed. The
symplectic condition for $S,$ $S_{3}^{T}S_{1}=S_{1}^{T}S_{3}$ brings
$(n^{2}-n)/2$ constrains on $w$ and $\dot{w}$, and the $O(n)$
gauge freedom for the equivalent class is also $(n^{2}-n)/2$. The
dimension of the decomposition is therefore
\begin{equation}
2n^{2}+n^{2}-(\frac{n^{2}-n}{2}+\frac{n^{2}-n}{2})=n(2n+1)\,,
\end{equation}
the same as the dimension of $Sp(2n).$ 

According to the polar decomposition theorem, any non-degenerate square
matrix $X$ can be uniquely factored into an orthogonal matrix $O$
and a symmetric, positive-definite matrix $Q,$ i.e., $X=OQ.$ As
a matter of fact, $Q=\sqrt{X^{T}X}$ and $O=XQ^{-1}.$ Using this
fact, at a fixed time $t=t_{1}$ we can always choose $c=\sqrt{w^{T}(t_{1})w(t_{1})}w^{-1}(t_{1})$
such that $\tilde{w}(t_{1})=cw(t_{1})$ is symmetric. With this gauge,
the canonical coordinate transformation at $t=t_{1}$ becomes
\begin{equation}
G=\tilde{P}\left(\begin{array}{cc}
\tilde{w}^{-1} & 0\\
(\tilde{w}R-\dot{\tilde{w}})m & \tilde{w}
\end{array}\right),
\end{equation}
which is in the form of a modified Iwasawa decomposition (or pre-Iwasawa
decomposition), whose importance in phase space optics \cite{Simon98,Wolf04-173}
and phase space quantum mechanics \cite{deGosson06-42} have been
recently realized. The modified Iwasawa decomposition is the unique
decomposition of a $2n\times2n$ symplectic matrix $G$ in the form
of 
\begin{equation}
G=P\left(\begin{array}{cc}
Y & 0\\
QY & Y^{-1}
\end{array}\right)\,,
\end{equation}
where $P\in Sp(2n)\bigcap SO(2n)=U(n)$ and $Y$ is symmetric. Matrix
$Q$ is also symmetric, which is equivalent to the condition $Y^{T}QY=(QY)^{T}Y$
for $\left(\begin{array}{cc}
Y & 0\\
QY & Y^{-1}
\end{array}\right)$ to be symplectic. These facts are also true if the decomposition
is alternatively defined to be
\begin{equation}
G=\left(\begin{array}{cc}
Y & 0\\
QY & Y^{-1}
\end{array}\right)P\,.
\end{equation}
For a symplectic matrix, the modified Iwasawa decomposition is equivalent
to the well-known Iwasawa decomposition for a semi-simple Lie group
\cite{Iwasawa49}. 

Making $\tilde{w}$ symmetric at $t=t_{1}$ fixes the $O(n)$ gauge
because $\tilde{w}(t_{1})$ is unique according to the polar decomposition
theorem. However, such a choice only makes $\tilde{w}$ symmetric
at $t=t_{1}.$ As in general gauge theories, we would like to pick
a gauge such that the envelope matrix is symmetric for all $t.$ To
accommodate this desired symmetry, we need to modify the governing
equations, especially the envelope equation. Let
\begin{align*}
w(t) & =c^{-1}(t)u(t)\,,\\
u(t) & =\sqrt{\beta(t)}=\sqrt{w^{T}(t)w(t)}\,,\\
c^{-1}(t) & =w(t)u^{-1}(t)\,,
\end{align*}
be the time-dependent polar decomposition of $w(t).$ Here we use
$u(t)$ to denote this special $\tilde{w}(t)$, which is symmetric
and positive-definite for all $t.$ The matrix $u(t)$ is the ``symmetrized''
$w(t)$ and equals the square-root of the generalized $\beta$ function.
We will recast the envelope equation \eqref{w} in terms of $\beta$,
as in the original Courant-Snyder theory for one degree of freedom
\cite{Courant58}. The difference is that the procedure here has to
be carried out in matrix form. 

Rewrite Eq.\,\eqref{w} as 
\begin{align}
\ddot{w} & +\dot{w}g+wh=w^{-T}m^{-1}w^{-1}w^{-T}m^{-1}\,,\label{eq:wgh}\\
g & \equiv(\dot{m}-Rm+mR^{T})m^{-1}\,,\\
h & \equiv(\kappa-RmR^{T}-\dot{R}m-R\dot{m})m^{-1}\,.
\end{align}
We symmetrize Eq.\,\eqref{eq:wgh} by taking $w^{T}($Eq.\,\eqref{eq:wgh}$)+($Eq.\,\eqref{eq:wgh}$)^{T}w$
to obtain a second order ordinary differential equation for $\beta,$
\begin{equation}
\ddot{\beta}-2\dot{w}^{T}\dot{w}+w^{T}\dot{w}g+g^{T}\dot{w}^{T}w+\beta h+h^{T}\beta=2m^{-1}\beta^{-1}m^{-1}\,.\label{eq:b''}
\end{equation}
It is a second-order equation for $\beta$ because $\dot{w}^{T}\dot{w}$
and $w^{T}\dot{w}$ can be expressed in terms of $\beta$ and $\dot{\beta}$
as follows. First note that
\begin{align}
\dot{w}^{T}\dot{w} & =\dot{u}Du+\dot{u}^{2}-uD^{2}u-uD\dot{u}\,,\label{eq:wtw}\\
w^{T}\dot{w} & =uDu+u\dot{u}\,,\label{eq:wtwd}\\
D & \equiv-\dot{c}c^{-1}\,.
\end{align}
Both $\dot{u}$ and $D$ in Eqs.\,\eqref{eq:wtw} and \eqref{eq:wtwd}
can be expressed as functions of $\beta$ and $\dot{\beta}.$ For
$\dot{u},$ from the definition of $u$ we obtain
\begin{equation}
\dot{u}u+u\dot{u}=\dot{\beta}\,,
\end{equation}
whose left-hand side can be viewed as a linear operator on $\dot{u}$
associated with $u$, 
\begin{equation}
L_{u}(\dot{u})\equiv\dot{u}u+u\dot{u}\,.\label{eq:Lu}
\end{equation}
The properties of the linear operator $L$ is discussed in the Appendix.
Since $u=\sqrt{\beta}$ is symmetric and positive-definite, $L_{u}$
is invertible to give 
\begin{equation}
\dot{u}=L_{\sqrt{\beta}}^{-1}(\dot{\beta})\,,
\end{equation}
where $L^{-1}$ is the inverse of $L$ defined in Eq.\,\eqref{eq:L-1}. 

To express $D$ in terms of $\beta$ and $\dot{\beta}$, we exam the
symplectic condition 
\begin{align}
wS_{3}^{T} & =S_{3}w^{T}\,,\\
S_{3} & \equiv-\dot{w}m+wRm\,.
\end{align}
Substituting in the polar decomposition $w=c^{-1}u$ gives
\begin{equation}
L_{umu}(D)=umuD+Dumu=(um\dot{u}-\dot{u}mu)+u(Rm-mR^{T})u\,.\label{eq:LD}
\end{equation}
Therefore, 
\begin{equation}
D=L_{umu}^{-1}\left[(um\dot{u}-\dot{u}mu)+u(Rm-mR^{T})u\right]\,,
\end{equation}
where $u=\sqrt{\beta}$ and $\dot{u}=L_{u}^{-1}(\dot{\beta})\,$.

Equation \eqref{eq:b''} is a second equation for $\beta$. Its solutions
do not uniquely determine the envelope matrix $w$, which is not surprising
considering that $\beta=w^{T}w$ is a ``symmetric'' version of $w$.
However, due to the $O(n)$ gauge freedom, $\beta$ contains enough
information to determine the transfer map $M.$ In terms of $u$ and
$c^{-1},$ 
\begin{align}
S & =\left(\begin{array}{cc}
c^{-1} & 0\\
0 & c^{-1}
\end{array}\right)S_{u}\,,\\
S_{u} & \equiv\left(\begin{array}{cc}
u^{-1} & 0\\
(uR-Du-\dot{u})m & u
\end{array}\right)\,.\label{eq:Su}
\end{align}
Even though the rotation matrix $c(t)$ here is a function of $t$,
the transformed phase advance is defined the same way as in the case
of a global gauge, i.e., 
\begin{equation}
P_{u}=P\left(\begin{array}{cc}
c^{-1} & 0\\
0 & c^{-1}
\end{array}\right)\,.\label{eq:Pud}
\end{equation}
What is modified is the governing equation for $P_{u}$, 
\begin{align}
\dot{P}_{u} & =-P_{u}\left[\left(\begin{array}{cc}
0 & \mu_{u}\\
-\mu_{u} & 0
\end{array}\right)-\left(\begin{array}{cc}
D & 0\\
0 & D
\end{array}\right)\right]\,,\label{eq:Pu}\\
\mu_{u} & \equiv\left(umu^{T}\right)^{-1}\,.
\end{align}
The second term on the right-hand side of Eq.\,\eqref{eq:Pu} is
due to the dependence on $t$ of the local gauge. At last, the canonical
coordinate transformation between $z$ and $\bar{\bar{z}}$ is 
\begin{equation}
\bar{\bar{z}}=Gz=P_{u}S_{u}z\,,\label{eq:SuPu}
\end{equation}
and the transfer map is 
\begin{equation}
M(t_{0},t)=S_{u}^{-1}P_{u}^{-1}S_{u0}\,.\label{eq:Ms}
\end{equation}
The symmetric decomposition furnished by Eqs.\,\eqref{eq:Su}, \eqref{eq:Pud},
and \eqref{eq:Ms} is equivalent to the decomposition described in
Sec. \ref{sec:Generalized-Courant-Snyder}, but it has three desirable
features by comparison. The canonical coordinate transformation $P_{u}S_{u}$
in Eq.\,\eqref{eq:SuPu} has the modified Iwasawa format for all
$t$. It comprises a curve of the modified Iwasawa decomposition,
developed from a dynamical point of view. The gauge freedom is removed,
and the dimension of the symplectic transfer map is directly reflected
by the dimension of the decomposition. At every $t$, $M(t)$ is specified
by two $n\times n$ symmetric matrices $\beta$ and $\dot{\beta},$
and a $U(n)$ matrix $P_{u}.$ The dimension of $M(t)$ is thus $(n^{2}+n)/2+(n^{2}+n)/2+n^{2}=n(2n+1)$. 

Before ending this section, we emphasize that the purpose of studying
the gauge freedom is to simplify the calculation of the symplectic
map $M$ and other lattice functions and beam parameters. By investigating
the gauge freedom in the matrix envelope equation for $w$, we have
found that we can actually bypass this gauge freedom and solve for
the $\beta$ matrix instead, which is symmetric and does not have
the gauge freedom. From Eqs.\,\eqref{eq:alphag}, \eqref{eq:gammag}
and \eqref{bg}, the generalized Twiss parameters $\alpha$ and $\gamma$
can also be expressed in terms of $\beta$ and $\dot{\beta}$. One
important advantage of using the $\beta$ matrix is that the symmetric
matrices $\beta$ and $\dot{\beta}$ form a linear space, which makes
the numerical algorithms of searching for matched solutions for $\beta$
much more efficient than for matched solutions for $w.$

\section{Stability analysis -- spectral stability and structural stability\label{sec:Stability-analysis}}

The classical analysis by Courant and Snyder \cite{Courant58} on
the instability induced by sum resonance for uncoupled transverse
dynamics may give a wrong impression that coupling effects are always
deleterious. The coupled dynamics can be stable or unstable depending
on the specific configuration of the lattice, but certainly not more
unstable than the uncoupled dynamics. The parameter space for a stable
coupled lattice is probably much larger than that of a stable uncoupled
lattice. In the conceptual design of the M\"{o}bius accelerator \cite{Talman95}
and N-rolling lattice \cite{Qin11-056708,Chung13}, it was argued
that strongly coupled lattice are more preferable for high-intensity
beams. Strongly coupled systems have been implemented in the spiral
line induction accelerator (SLIA) \cite{Gluckstern79,Roberson1985,Chernin1988,Petillo1989,Krall1995,Smith1998,Petillo2002},
which reached up to 10KA electron current at 5 MeV beam energy. Our
understanding of the stability properties of coupled dynamics has
been limited by the theoretical tools available. In this section,
we demonstrate how the generalized Courant-Snyder theory can be applied
to study the stability of the most general focusing lattice given
by Eq.\,\eqref{H} with weak and strong coupling components in realistic
accelerators. 

For a thorough understanding, it is necessary to distinguish two types
of linear stability (or instability). The first type is spectral stability,
which means the linear dynamics is stable for all initial perturbations.
The system is spectrally unstable if there exists an initial condition
that grows without bond. In most contexts, the meaning of stability
is that of spectral stability. The second type is the so-called structural
stability (or strong stability). It mostly applies to systems that
are spectrally stable. A spectrally stable system is structurally
unstable if there is a spectrally unstable system infinitesimally
closed-by. Otherwise, the spectrally stable system is also structurally
stable. The well-known result with respect to the stability properties
of sum/difference resonances for uncoupled lattices refers to the
structural stability under the influence of an infinitesimal coupling
component \cite{Courant58}. 

The spectral and structural stability of the transverse dynamics in
a periodic focusing lattice is determined by its one-turn (or one-period)
map $M(T)$. The fact that $M(T)$ is a symplectic matrix regulates
the stability properties in a significant way \cite{Krein50,Gelfand55,Moser58,Yakubovich75,DragtBook}.
We list here the relevant results without presenting details of the
proof. 

The spectral property is determined by the eigenvalues and their multiplicities.
There are four possibilities:
\begin{lyxlist}{00.00.0000}
\item [{C1)}] All eigenvalues are distinct and on the unit circle of the
complex plane.
\item [{C2)}] All eigenvalues are on the unit circle. There are repeated
eigenvalues. But the geometric multiplicity for all eigenvalues is
the same as the algebraic multiplicity.
\item [{C3)}] All eigenvalues are on the unit circle. There are repeated
eigenvalues with algebraic multiplicity greater than the geometric
multiplicity.
\item [{C4)}] There exits at least one eigenvalue not on the unit circle.
\end{lyxlist}
Cases C3 and C4 are spectrally unstable, and Cases C1 and C2 are spectrally
stable. For Cases C1 and C2, we would like to know whether they are
also structurally stable. It has been shown that Case C1 is structurally
stable using the symplectic nature of $M(T)$ \cite{Krein50,Gelfand55,Moser58,Yakubovich75,DragtBook}.
Case C2 needs to be sub-divided into two categories:
\begin{lyxlist}{00.00.0000}
\item [{C2.1)}] For all repeated eigenvalues, the corresponding eigenvectors
have the same signatures. 
\item [{C2.2)}] There is at least one repeated eigenvalue whose eigenvectos
have different signatures. 
\end{lyxlist}
According to the Krein-Moser theorem \cite{Krein50,Gelfand55,Moser58,Yakubovich75,DragtBook},
Case C2.1 is structurally stable and C2.2 is structurally unstable.
For an eigenvector $\psi$ of $M(T),$ its signature is defined to
be the sign of its self-product $\left\langle \psi,\psi\right\rangle =\psi^{*}iJ\psi.$
The product between two eigenvectors $\psi$ and $\phi$ in general
is defined to be $\left\langle \psi,\phi\right\rangle \equiv\psi^{*}iJ\phi,$
where $\psi^{*}$denotes the complex conjugate of $\psi^{T}.$

To design a coupled lattice, it is desirable to be in Case C1, which
is both spectrally and structurally stable. As mentioned previously,
for the general Hamiltonian given by Eq.\,\eqref{H}, the parameter
space satisfying this condition is large enough for most applications.
Given a periodic lattice, we can search for a matched solution for
$\beta,$ as in the original Courant-Snyder theory for one degree
of freedom \cite{Courant58}. After a matched $\beta$ is found, the
one-turn map is 
\[
M(T)=S_{0}^{-1}P(T)^{-1}S_{0}\,,
\]
which implies that $M(T)$ is similar to $P(T)^{-1}$. Their eigenvalues
and multiplicity are identical. Because $P(T)$ is a symplectic rotation,
all of its eigenvalues are on the unit circle, automatically ruling
out the unstable situation in C4. 

The phase advance $P(T)$ also determines the structural stability
of the system. To prove this assertion, let $\psi$ and $\phi$ are
the eigenvectors of $M(T)$. Then $S_{0}\psi$ and $S_{0}\phi$ are
the eigenvectors of $P(T)^{-1},$ and 
\begin{equation}
\left\langle S_{0}\psi,S_{0}\phi\right\rangle =\psi^{*}S_{0}^{T}iJS_{0}\phi=\psi^{*}iJ\phi=\left\langle \psi,\phi\right\rangle \,,\label{eq:spsiphi}
\end{equation}
where use had been made of the fact $S_{0}$is symmetric, i.e., $S_{0}^{T}JS_{0}=J.$
Equation \eqref{eq:spsiphi} states that the signatures of eigenvectors
of $P(T)^{-1}$ and thus its structural stability are identical to
that of $M(T).$

These analyses lead to the important conclusion that the phase advance
matrix $P(T)$ completely determines both the spectral and structural
stability of the general focusing lattices. This fact can significantly
simplify the stability analysis in lattice design. For example, if
the system is in Case C2, we only need to look at the signatures of
the eigenvectors of $P(T)^{-1}$ to know if it is structurally stable.
According the Krein-Moser theorem, if the eigenvectors for all repeated
eigenvalues of $P(T)^{-1}$ have the same signatures, then the system
is structurally stable. Otherwise, it is structurally unstable. Let's
show that this conclusion recovers the classical results on the stability
properties of sum/difference resonances for uncoupled quadrupole lattices
as special cases. In this case, the phase advance matrix is calculated
to be \cite{Qin09-NA,Qin09PoP-NA} 
\[
P(T)^{-1}=\left(\begin{array}{cccc}
\cos\phi_{x} & 0 & \sin\phi_{x} & 0\\
0 & \cos\phi_{y} & 0 & \sin\phi_{y}\\
-\sin\phi_{x} & 0 & \cos\phi_{x} & 0\\
0 & -\sin\phi_{y} & 0 & \cos\phi_{y}
\end{array}\right)\,,
\]
where $\phi_{x}$ and $\phi_{y}$ are the one-turn phase advance in
the $x-$ and $y-$directions. Its four sets of eigenvalues, eigenvectors,
and signatures are
\begin{align}
\lambda_{x+} & =\cos\phi_{x}+i\sin\phi_{x}\,,\,\,\,\psi_{x+}=(1,0,i,0)^{T}\,,\,\,\,\sigma_{x+}=-1\,,\\
\lambda_{x-} & =\cos\phi_{x}-i\sin\phi_{x}\,,\,\,\,\psi_{x-}=(1,0,-i,0)^{T}\,,\,\,\,\sigma_{x-}=1\,,\\
\lambda_{y+} & =\cos\phi_{y}+i\sin\phi_{y}\,,\,\,\,\psi_{y+}=(0,1,0,i)^{T}\,,\,\,\,\sigma_{y+}=-1\,,\\
\lambda_{y-} & =\cos\phi_{y}-i\sin\phi_{y}\,,\,\,\,\psi_{y-}=(0,1,0,-i)^{T}\,,\,\,\,\sigma_{y-}=1\,.
\end{align}
Resonance occurs when two or more eigenvalues collide, which has four
possibilities:
\begin{lyxlist}{00.00.0000}
\item [{R1)}] Self-resonance in the $x-$direction. $\phi_{x}=n\pi$ and
$\lambda_{x+}=\lambda_{x-}=\pm1.$
\item [{R2)}] Self-resonance in the $y-$direction. $\phi_{y}=n\pi$ and
$\lambda_{y+}=\lambda_{y-}=\pm1.$
\item [{R3)}] Sum resonance. $\phi_{x}+\phi_{y}=n\pi,$ $\lambda_{x+}=\lambda_{y-}$
and $\lambda_{x-}=\lambda_{y+}.$
\item [{R4)}] Difference resonance. $\phi_{x}-\phi_{y}=n\pi,$ $\lambda_{x+}=\lambda_{y+}$,
and $\lambda_{x-}=\lambda_{y-}.$
\end{lyxlist}
Case R1 is structurally unstable because $\sigma_{x+}$ and $\sigma_{x-}$
are different. Case R2 is structurally unstable for the same reason. 

For the sum resonance at the repeated eigenvalue $\lambda_{x+}=\lambda_{y-}$,
the signatures $\sigma_{x+}$ and $\sigma_{y-}$ of the corresponding
eigenvectors have different signs. The sum resonance is thus structurally
unstable. For the difference resonance at the first repeated eigenvalue
$\lambda_{x+}=\lambda_{y+}$, the corresponding eigenvectors $\psi_{x+}$
and $\psi_{y+}$ have the same signature. This is also true at the
second repeated eigenvalue $\lambda_{x-}=\lambda_{y-}.$ The difference
resonance is thus structurally stable. These results are well-known
previously \cite{Courant58}, but recovered here as a special case
of a more general criterion based on the generalized phase advance
and the Krein-Moser theorem \cite{Krein50,Gelfand55,Moser58,Yakubovich75,DragtBook}.
We expect that the more general stability criterion expressed in terms
of the phase advance matrix $P(T)$ to be a powerful tool for future
lattice design with strong coupling.

\section{Conclusions and future work\label{sec:Conclusions}}

We have presented in this paper a detailed derivation of the generalized
Courant-Snyder theory for the most general linear focusing lattices
with quadrupole, skew-quadrupole, dipole, and solenoidal components,
as well as torsion of the fiducial orbit and variation of beam energy.
The theoretical structure of the theory in terms of gauge freedoms
and group decomposition were described. We have also demonstrated
the application of the theory in stability analysis for strongly and
weakly coupled lattices. In addition to being more realistic, the
most general Hamiltonian in Eq.\,\eqref{H} enables a much larger
parameter space for designing strongly coupled lattices that are spectrally
and structurally stable. The generalized Courant-Snyder parameterization
scheme developed here provides an effective tool to study the coupled
dynamics and to discover more optimized lattice design in the larger
parameter space of general focusing lattices. The formalism also sets
the theoretical foundation for investigating collective phenomena
in high-intensity beams, such as the self-consistent solutions of
the Vlasov-Maxwell equations in phase space including strong self-field
effects that can couple the transverse dynamics \cite{Wang82,Chao93-all,Davidson01-284,Lund04,Zimmermann04-124801}. 

As mentioned in Sec. \ref{sec:Generalized-Courant-Snyder}, the theoretical
framework developed is valid for linear system with any degree of
freedom. In particular, we can apply it to the 3D coupled dynamics,
which includes the sychrotron oscillation in RF cavities, and the
linear coupling between transverse and longitudinal dynamics as in
the recent investigations of emittance change \cite{Cornacchia02,Emma06,Qin11-758}.
In this case, $n=3$ and the focusing matrix $\kappa(t)$ in Eq.\,\eqref{H}
becomes a $3\times3$ matrix which describes both sychrotron and betatron
oscillations as well as possible coupling between them. The envelop
matrix $w$ is $3\times3$ and satisfies Eq.\,\eqref{w} with $R$,
$m$ and $\kappa$ being $3\times3$ matrices. The Twiss parameters
$\alpha,$ $\beta$, and $\gamma$ are $3\times3$ matrices, the symplectic
matrix $M$ is $6\times6$, and all the equations they satisfy are
the same as in the case of 2-degrees of freedom. Studies in these
directions will be reported in future publications. 

\appendix*

\section{$L_{x}(y)$}

In this Appendix, we derive the mathematical properties of the linear
transformation used in Eqs.\,\eqref{eq:Lu} and \eqref{eq:LD}. Let
$A$ and $X$ denote $n\times n$ matrices. For a symmetric, positive-definite
matrix $A$, define the linear function associated with $A$ on $X$
as 
\[
L_{A}(X)\equiv AX+XA\,.
\]
We prove that $L_{A}$ is invertible. It is enough to show that $L_{A}$
is injective, i.e., $L_{A}(X)=0$ only if $X=0$. Let $X$ is in the
kernel of $L_{A},$ i.e., 
\begin{equation}
L_{A}(X)=0.\label{eq:LA}
\end{equation}
Since $A$ is symmetric, the eigenvectors of $A$ form a basis for
vectors in $R^{n}.$ Expressed in this basis, Eq.\,\eqref{eq:LA}
is 
\begin{equation}
\lambda_{u}v^{T}Xu+\lambda_{v}v^{T}Xu=0\,,\label{eq:lulv}
\end{equation}
where $u$ and $v$ are any pairs of vectors in the basis, and $\lambda_{u}$
and $\lambda_{v}$ are the corresponding eigenvalues, respectively.
Because the eigenvalues of a positive-definite matrix are positive,
Eq.\,\eqref{eq:lulv} is possible only when $v^{T}Xu=0$. This proves
that $L_{A}$ is invertible. In terms of its components in this basis,
$L_{A}^{-1}$ is given as 
\begin{equation}
u^{T}Xv=\frac{u^{T}Yv}{\lambda_{u}+\lambda_{v}}\,,\label{eq:L-1}
\end{equation}
where $Y=L_{A}(X).$
\begin{acknowledgments}
This research was supported by the U.S. Department of Energy (DE-AC02-09CH11466
and DE-AC02-07CH11359). 
\end{acknowledgments}

%\bibliographystyle{apsrev}
%\bibliography{../Refs}

\end{document}